\documentclass[aps]{revtex4}
\pdfoutput=1 
\usepackage[T1]{fontenc}
\usepackage[latin9]{inputenc}
\usepackage{amsmath}
\usepackage{graphicx}
\usepackage{amssymb}
\usepackage{setspace}
\usepackage{color}
\usepackage{fancyhdr}

\begin{document}
\title{On the Infrared Behavior of Landau Gauge Yang-Mills Theory \\with a Fundamentally Charged Scalar Field}
\author{Leonard Fister$^{1,2,3}$, Reinhard Alkofer$^{1}$ and Kai Schwenzer$^{1,4}$}
\affiliation{$^{1}$Institut f\"ur Physik, Karl-Franzens-Universit\"at Graz, Universit\"atsplatz
5, 8010 Graz, Austria}
\affiliation{$^2$Institut f\"ur Theoretische Physik, Ruprecht-Karls-Universit\"at Heidelberg, Philosophenweg 16, 69120 Heidelberg, Germany}
\affiliation{$^3$ExtreMe Matter Institute EMMI, GSI Helmholtzzentrum f\"ur Schwerionenforschung, Planckstr. 1, 64291 Darmstadt, Germany}
\affiliation{$^{4}$Department of Physics, Washington University, St. Louis, MO
63130, USA}

\begin{abstract}
Recently it has been shown that infrared singularities of Landau gauge QCD
can confine static quarks via a linearly rising potential. We show that the
same mechanism can also provide a confining interaction between charged scalar fields in the fundamental representation. This confirms
that within this scenario static confinement is a universal property
of the gauge sector even though it is formally represented in the
functional equations of the matter sector. The simplifications compared
to the fermionic case make the scalar system an ideal laboratory for
a detailed analysis of the confinement mechanism in numerical studies
of the functional equations as well as in gauge-fixed lattice simulations.
\end{abstract}
\maketitle

\section{Introduction}
\thispagestyle{fancy}
The confinement of quarks is a remarkable phenomenon that
features both static and dynamical aspects. To reveal its underlying
mechanism it is on the one hand essential to restrict the analysis
by considering certain limits where the system is considerably simpler
and particular aspects of the mechanism become transparent. On
the other hand one extends the analysis to similar systems, in order
to identify the requirements and characteristic properties of the
mechanism. An important limit in this respect is the static case and the corresponding confinement of
fundamental color sources which presents an analytic
result of non-Abelian gauge theory in the strong-coupling limit \cite{Wilson:1974sk}
and which has been confirmed in numerical simulations at realistic
coupling. This is reflected by an area law behavior of large Wilson
loops corresponding to a linear potential between the sources and presents a genuine property
of the gauge dynamics that is independent of the detailed aspects
of the sources and merely depends on the representation of the
gauge group. \\
Within the last years a detailed picture of the infrared (IR) sector of Yang-Mills theory and QCD has evolved, mainly due to investigations within either functional approaches as Dyson-Schwinger equations \cite{vonSmekal:1997is,Fischer:2006ub,Alkofer:2000wg,Watson:2001yv,Lerche:2002ep,Zwanziger:2001kw,Zwanziger:2002ia,Alkofer:2008tt,Schwenzer:2008vt,Fischer:2002hna,Alkofer:2004it,Huber:2007kc,Fischer:2006vf,Fischer:2007pf,Fischer:2008uz,Alkofer:2008jy,Huber:2009wh,Huber:2009tx} and functional renormalisation group equations \cite{Fischer:2008uz,Litim:1998nf,Pawlowski:2003hq,Gies:2006wv,Fischer:2004uk,Pawlowski:2005xe}, respectively, or lattice gauge theory \cite{Cucchieri:2008fc,Sternbeck:2008mv,vonSmekal:2008ws}. 
It has been shown that the infrared singularities of Landau gauge quantum chromodynamics (QCD) can
provide a mechanism for the confinement of static quarks \cite{Alkofer:2008tt}.
This mechanism, relying on the IR scaling solution of Yang-Mills theory
\cite{vonSmekal:1997is,Fischer:2006vf}, is driven by a strong kinematic singularity
of the quark-gluon vertex that overturns the IR suppression of the
gluon propagator and leads to a long-range gluonic interaction. Yet,
in this approach this is inherently obtained from the static limit
of the solution of the functional equations of the quark sector of
the theory. Since the mechanism exhibits a relation between chiral
symmetry breaking and confinement, this poses the question whether
the Dirac structure of the quarks is essential for this mechanism.
In order to answer this question we present here results for a related theory where
the quarks are replaced by a fundamentally charged scalar field, for details we refer to \cite{Leo}. This
theory is interesting because of its simplicity owing to the absence
of internal spin degrees of freedom compared to the fermionic
theory and has been studied before as a model system for the QCD dynamics.
On the other hand, this theory involves additional self-interactions
between the scalars and thereby could exhibit a rather different dynamical
behavior. Most notably it involves a Higgs phase in addition to a
confining phase, and it has been shown that in the fundamentally charged scalar model the confined
and the Higgs phase are not separated by a phase boundary \cite{Osterwalder:1977pc}. \\
In dynamical QCD a gluonic interaction that rises with distance is only realized over
a certain range that varies continuously with the quark mass(es) \cite{Schwenzer:2008vt}.
A simplified model system to study all these aspects would be highly
desirable. Interestingly string breaking signatures have likewise
been observed in lattice simulations of the scalar model, cf. e.g.
\cite{Bock:1988kq}, before corresponding studies were possible in
dynamical QCD, cf. e.g. \cite{Bali:2005fu}.

\section{Dynamics of fundamentally charged scalar fields \break coupled to Landau gauge Yang-Mills Theory}
\begin{figure}
\parbox[c][1\totalheight]{0.5\columnwidth}{
\includegraphics[scale=0.17]{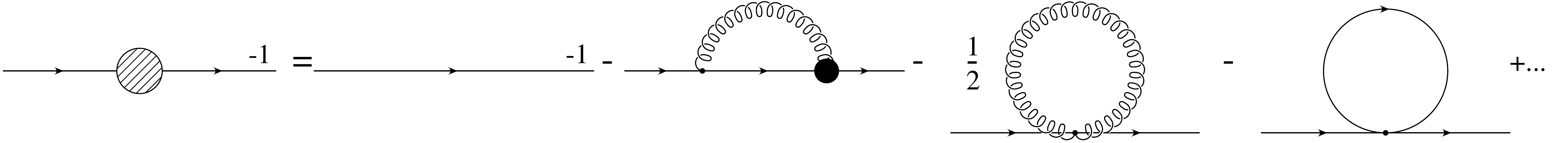}\\
 \vspace*{0.4cm}\includegraphics[scale=0.17]{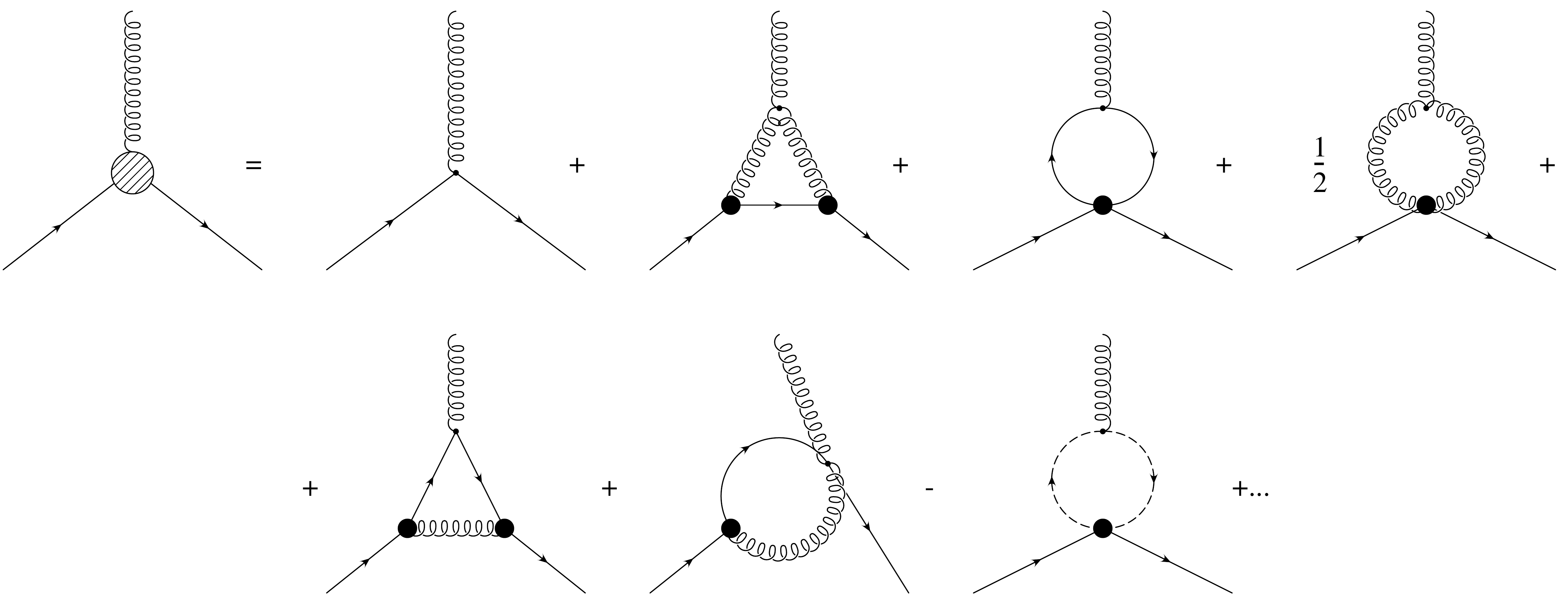}\\
 \vspace*{0.4cm}\includegraphics[scale=0.17]{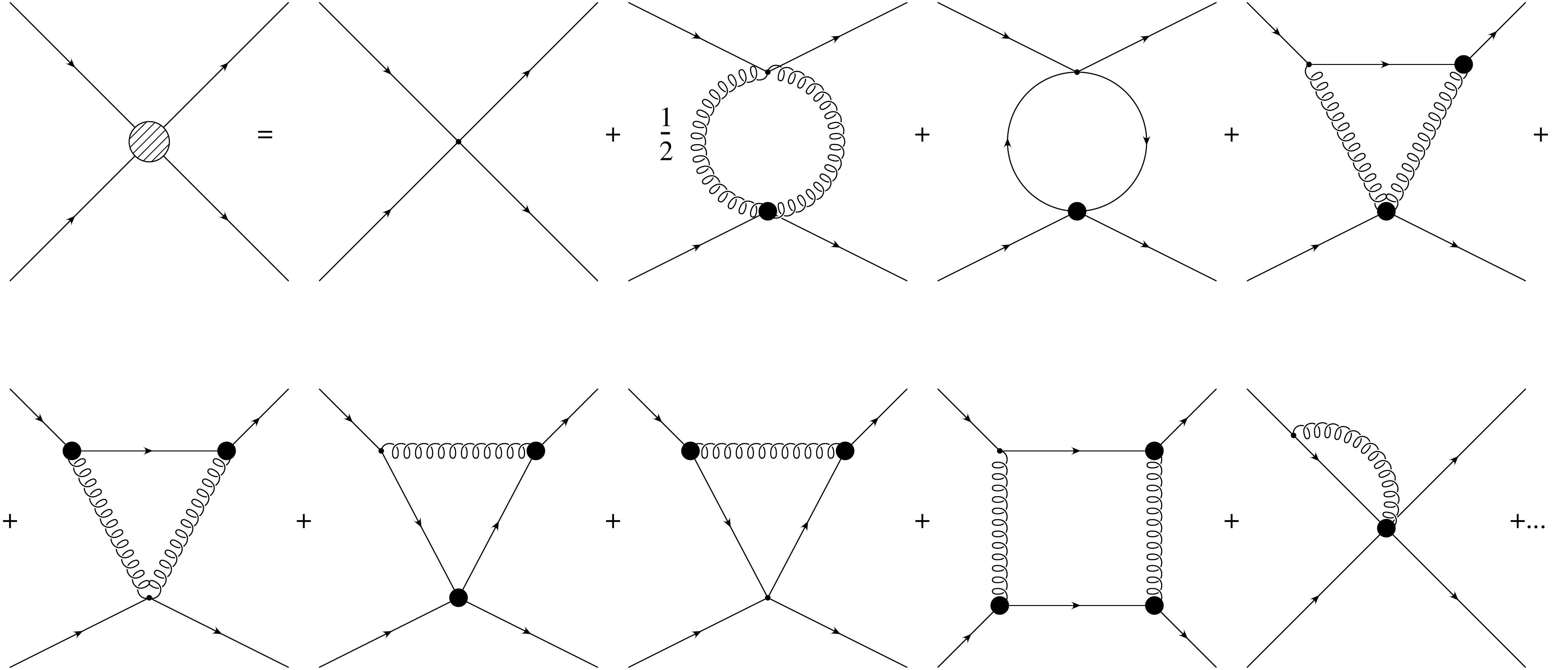}
}\begin{minipage}[c][1\totalheight]{0.5\columnwidth}%
\flushright\includegraphics[scale=0.17]{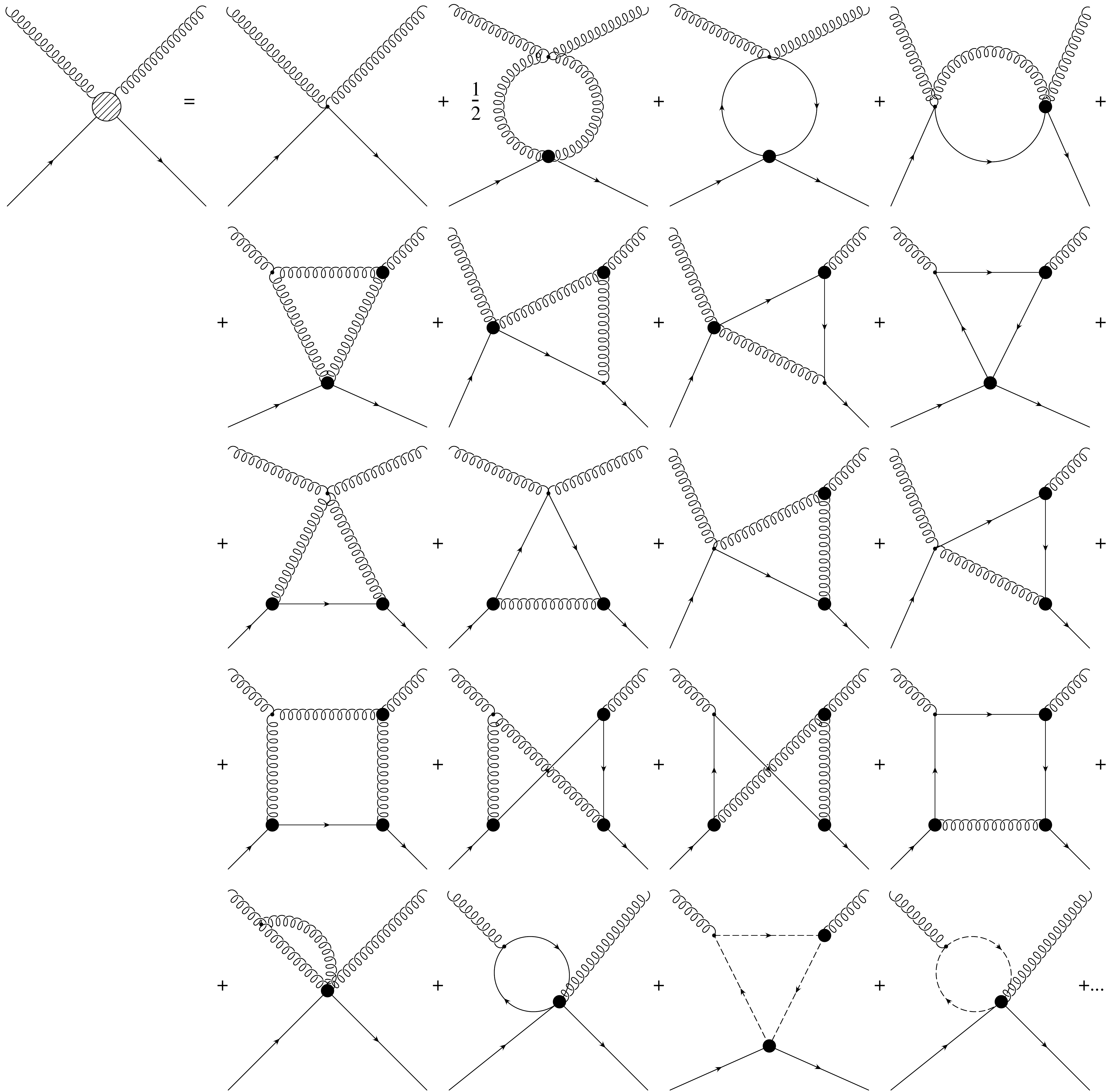}%
\end{minipage}
\caption{The matter part of the considered DSE system.\label{fig:scalar-DSEs}\emph{
left}: The DSEs for the scalar propagator, the scalar-gluon vertex
and the 4-scalar vertex; \emph{right}: The equation for the scalar-2-gluon
vertex. All 2-loop contributions in these equations are IR suppressed. Permutations of the given diagrams and two-loop diagrams are denoted by the ellipses.}
\end{figure}

\begin{figure}
\parbox[c][1\totalheight]{0.5\columnwidth}{ \includegraphics[scale=0.1]{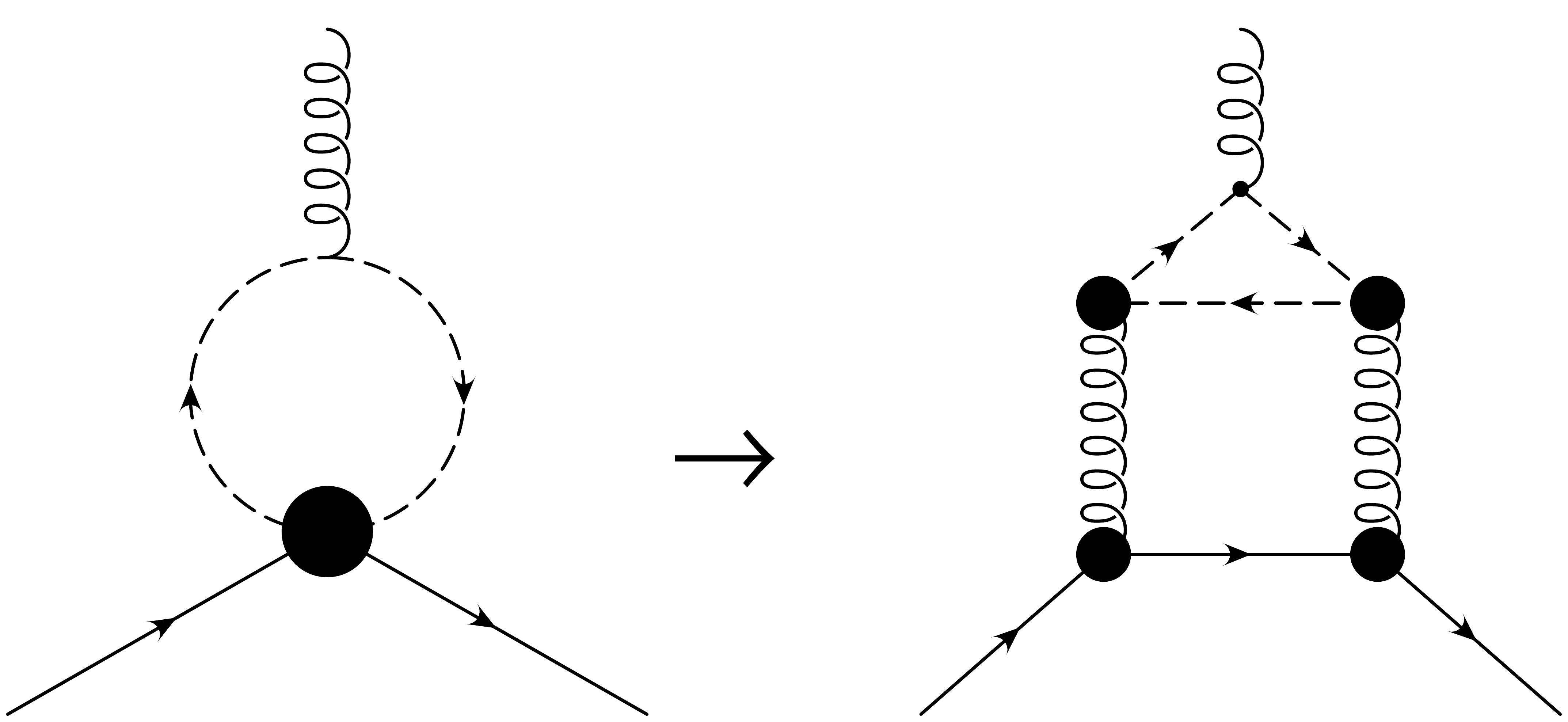}
}\begin{minipage}[c][1\totalheight]{0.5\columnwidth}
\flushright \includegraphics[scale=0.17]{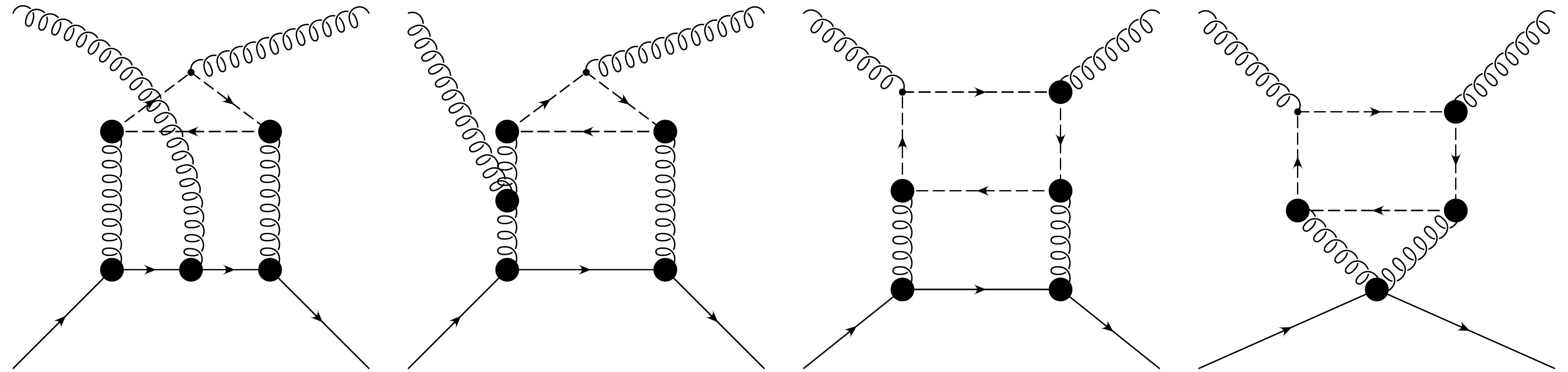}
\end{minipage}
\caption{\label{fig:gh-box}\emph{left:} The lowest order ghost contribution to the scalar-gluon vertex in the skeleton expansion. \emph{right:} By the same mechanism in the 2-scalar-2-gluon vertex equation analogous diagrams arise, stemming from the last two diagrams (the ghost-triangle and the ghost-loop diagram) of fig. \ref{fig:scalar-DSEs}.}
\end{figure}

In order to construct a model system for full QCD, where (fermionic) quarks carry a conserved charge and transform according to the fundamental representation of the gauge group, we attribute the same transformation properties also to the (bosonic) scalars. Therefore the scalars will firstly be implemented in the fundamental representation, and secondly the scalar field must be a complex field in order to be able to define a conserved charge. These scalars will be coupled to an $SU\!\left(N\right)$ gauge theory, and as we aim at comparing the system to Landau gauge QCD we also fix to Landau gauge, i.e. we take the limit for the gauge fixing parameter $\zeta \rightarrow 0 $.
Considering only renormalizable interactions this generally results in the Lagrangian given by
\begin{equation}
\mathcal{L}=\left(D_{\mu,ij}\phi_{j}^{*}\right)\left(D_{\mu,ik}\phi_{k}\right)-m^{2}\phi_{i}^{*}\phi_{i}-\frac{\lambda}{4!}\left(\phi_{i}^{*}\phi_{i}\right)^{2}+\frac{1}{4}F_{\mu\nu}^{a}F_{\mu\nu}^{a}+\frac{1}{2\zeta}(\partial_{\mu}A_{\mu}^{a})^{2}+\bar{c}^{a}\partial_{\mu}D_{\mu}^{ab}c^{b},
\end{equation}
with
\begin{equation}
D^{ab}_{\mu} = \delta^{ab} \partial_{\mu} + g f^{abc}A_{\mu}^{c}, \ \ \ \ D_{\mu,ij} = \delta_{ij} \partial_{\mu} - i g \left(\frac{t^{a}}{2}\right)_{ij} A_{\mu}^{a}, \ \ \ \ F_{\mu \nu}^{a} = \partial_{\mu} A_{\nu}^{a} - \partial_{\nu} A_{\mu}^{a} - g f^{abc} A_{\mu}^{b} A_{\nu}^{c},
\end{equation}
wherein $D_{\mu,ij}$ denotes the covariant derivative, involving the Gell-Mann matrices $t^a$, for the complex scalar field $\phi^{(*)}$ with the associated mass $m$, $\lambda$ is the coupling constant for a quartic scalar interaction, $F_{\mu\nu}^{a}$ is the field-strength tensor involving the gluons $A$ and the structure constants $f^{abc}$, and $D_{\mu}^{ab}$ is the covariant derivative for the Faddeev-Popov (anti-)ghosts $(\bar{c})c$ in the adjoint representation. Lorentz-indices are written in Greek letters, roman indices starting with $a$ are color-indices and the fundamental representation is indexed by roman letters starting with $i$.\newline
In contrast to QCD the tensor structure of the scalar
model is strongly simplified. In the quark propagator one has to consider a Dirac-vector as well as a Dirac-scalar component. Instead for scalar bosons, there is only one (scalar) tensor component in the scalar propagator $S^{ij}$. Similarly the scalar-gluon vertex $\Gamma^{a,ij}_{\mu}$ depending on two independent momenta can be decomposed into two tensors, in contrast to 12 independent tensors in the quark-gluon vertex. This simplification becomes even more significant for higher correlation functions. We will choose the parametrization 
\begin{equation}
S^{ij}\left(p\right)=-\delta_{ij}\frac{\tilde{S}\left(p^{2}\right)}{p^{2}}\:, \ \ \  
\Gamma_{\mu}^{a,ij}\left(p_{s},p_{gl}\right)=ig\left(t^{a}\right)_{ij}\left(\tilde{\Gamma}_{s}\left(p_{s}^{2},p_{gl}^{2},p_{s}\cdot p_{gl}\right)p_{s,\mu}+\tilde{\Gamma}_{gl}\left(p_{s}^{2},p_{gl}^{2},p_{s}\cdot p_{gl}\right)p_{gl,\mu}\right),
\end{equation}
where the two independent momenta are conveniently chosen as the incoming scalar and the gluon momentum.
Note that besides the simplification in the tensor structures two additional subtleties arise that are not present in QCD. Firstly, the classical Lagrangian contains additional 4-particle-interactions involving scalars, whose analogous terms are not present in QCD due to dimensional reasons (renormalizability). Secondly, a scalar field theory gives rise to a possible scalar condensate via the Higgs mechanism. \\
Here a note is in order. The theory described above develops a more complicated phase structure compared to QCD. As the gauge is already fixed the residual symmetry that is broken must be a global one \cite{Caudy:2007sf}. Note that it is ascertained that these two different "phases" of the system are not separated by a phase boundary \cite{Osterwalder:1977pc}, but rather by a Kert\'esz line, i.e. they are continuously connected. In that sense the terminology of "phases" is not strictly correct, but as it is frequently used in the literature we will adapt this nomenclature. A suitable quantity to investigate the phase transition is an order parameter \cite{Langfeld:2002ic} of the form
\begin{equation}
Q=\left(\int d^{4}x\ \phi(x)\right)\left(\int d^{4}x\ \phi^{\dagger}(x)\right)\:.
\end{equation}
The relation of screening to confinement has been subject to other investigations \cite{Gaete:2009xf}. In this work we will not discuss the effects due to scalar condensation and concentrate on the confinement properties of the model - leaving a detailed analysis of a condensate for future work.\\
In the following we will use the functional Dyson-Schwinger equations (DSEs) to describe the non-perturbative dynamics of the theory. The DSE analysis we perform here
relies on the framework described in detail in \cite{Alkofer:2004it,Fischer:2006vf,Alkofer:2008jy}
and is similar to the analysis in QCD \cite{Alkofer:2008tt,Schwenzer:2008vt}.
The details on the derivation of the presented results can be
found in \cite{Leo}. The DSEs for the theory can be derived algorithmically \cite{Alkofer:2008nt}. The corresponding equations
in the gauge sector are given in \cite{Alkofer:2008jy}, which also hold for coupled scalars in the quenched approximation. In the case of a dynamical scalar these equations are extended by unquenching graphs with closed scalar loops analogous to the case of QCD. The leading
equations in the scalar sector are shown in fig. \ref{fig:scalar-DSEs}
where 2-loop diagrams arising in these equations are omitted. These
equations represent an infinite tower of coupled integral equations and in general would require
some truncation. In order to study interesting qualitative
aspects, like the confinement of static sources, it
is sufficient to study the long-range behavior of the theory. In momentum
space this is encoded in the IR regime of correlation functions, and as far as only the IR scaling laws are concerned a solution of the whole tower is actually possible. 
This has previously been achieved via the help of a skeleton expansion
\cite{Alkofer:2004it} and we will also employ this approach in this work.
The skeleton expansion presents a loop expansion
in terms of dressed Green functions and this way the equations for the
primitively divergent Green functions are transformed into a closed
system of equations. However, as argued recently the skeleton expansion is only a convenient tool and not mandatory since the IR scaling is strongly restricted and fully determined by the equations for the primitively divergent correlation
functions, see e.g. \cite{Huber:2009wh,Fischer:2006vf,Schwenzer:2008vt} and references therein. Since in the scaling solution of Yang-Mills theory \cite{vonSmekal:1997is} the ghost dynamics is strongly dominant in the IR regime and there is no direct coupling between scalars and ghosts in the Lagrangian, the first ghost corrections in the scalar sector arise from two loop diagrams in the skeleton expansion. Correspondingly, we have to consider these graphs in our analysis whereas otherwise higher loop graphs in the skeleton expansion must not be more divergent than the corresponding lower order graphs.
The resulting leading ghost contribution in the scalar-gluon vertex DSE is given in fig. \ref{fig:gh-box}. Similar diagrams emerge from the ghost contributions to the scalar-2-gluon vertex equation. \\
Besides masses, QCD has no inherent scales far below the MeV scale. Thus in the IR
regime $p\ll\Lambda_{QCD}$ all Green functions should exhibit a scaling
form in terms of the external momenta. In the case that all external momenta vanish uniformly ({\em uniform} limit) with a single external scale $p$ power law scaling solutions of the form
\begin{equation}
\Gamma\left(p\right)\sim\left(p^{2}\right)^{\chi+\delta}\:
\end{equation}
are a natural ansatz for Green functions. Herein the canonical exponent $\chi$ reflects the dimension of
the corresponding operator and the anomalous exponent $\delta$ describes the anomalous scaling induced by the dynamics. In the case
of the scalar propagator and scalar-gluon vertex defined above the
corresponding canonical dimensions are $\chi_{s}=-1$ and $\chi_{sg}=1/2$.
In the uniform limit there is effectively only a single scale, and
the loop integrals that are dominated by the poles of the integrand
have to scale with this external scale. The scaling of a given loop
graph can then be determined by a power counting analysis. Yet, when
there are additional mass scales present the loop integrals can also
be dominated by scales of the order of the mass which has to be considered in detail
in the power counting analysis \cite{Alkofer:2008jy}. For a massive particle the propagator can alternatively be parametrized by a mass dressing function $M$ which features a massive IR behavior for $m\!\equiv\! M\!\left(0\right)\!>\!0$ and contains in this case an additional hard scale
\begin{equation}
\frac{\tilde{S}\left(p^{2}\right)}{p^{2}}=\frac{1}{p^{2}+M^{2}\left(p^{2}\right)} \sim \left(p^{2}\right)^{-1+\eta}.
\end{equation}
Therefore the mass dressing function $M$ yields an additional exponent $\eta$ in the power counting, with $\eta=1$ for massive particles, and $\eta=0$ for massless particles, respectively.


\section{Uniform infrared scaling}
In this section we discuss the fundamentally charged scalar model in the limit where all external momenta of Green functions vanish uniformly. In order to find such infrared solutions we perform a power counting analysis. The goal of this analysis is to find the scaling behavior of Green functions under the assumption that all Green functions scale with some power of $p^2$ in the deep infrared. In this case we can simply count the exponents of the different dynamical contributions in the DSEs and determine a solvable system for the scaling exponent of each Green function. As $p\rightarrow 0$ the dominant term which determines the scaling of the corresponding Green function is the one with the most divergent power law and correspondingly the one with the minimal IR exponent. With this procedure and the help of the skeleton expansion we obtain a closed system for the IR exponents of the primitively divergent vertex functions. For an illustrative example of this counting procedure consider the scalar propagator given in fig. \ref{fig:scalar-DSEs} in the top left corner. On the left hand side the power law ansatz gives $(p^2)^{1-\delta_s}$ since it is the inverse of the propagator that arises in the DSE, giving an additional overall-sign. The right hand side involves four different diagrams, and the dominant term is correspondingly given by a minimum function of the individual exponents. The first diagram is simply the inverse bare propagator giving $(p^2)^{1-\eta}$. In the second diagram one has to count the momentum dependence of the loop integration $(p^2)^2$, a scalar-propagator $(p^2)^{-1+\delta_s}$, a gluon-propagator $(p^2)^{-1+\delta_g}$ and a scalar-gluon vertex $(p^2)^{\frac{1}{2}+\delta_{sg}}$. Performing the analogous power counting for the last two graphs and subtracting the canonical dimensions one obtains an equation for the anomalous dimension of the scalar-propagator
\begin{equation}
-\delta_{s}= \min\left( -\eta, \ \delta_{s}+\delta_{gl}+\delta_{sg}, \ \delta_s, \ \delta_g \right).
\end{equation}
Applying the same procedure for all skeleton expanded DSEs of the primitively divergent vertex functions of pure gauge theory in \cite{Alkofer:2008jy} and the scalar sector in fig. \ref{fig:scalar-DSEs} one obtains a closed system of equations for the IR exponents of the primitively divergent vertex functions. Due to the scalar self-interaction this system of equations for the anomalous IR exponents proves to be rather extensive in the scalar model and is explicitly given in \cite{Leo}. \\
The next task is then to find the leading infrared behavior of all Green functions for the different infrared fixed points as solution of this purely algebraic system. As discussed in detail in \cite{Leo} it turns out that it is a convenient starting point to consider the equations from the gauge sector first, wherein the ghost equation is the most convenient one to start with. Depending on the boundary condition of the DSE that determines if the bare term is present or absent in the corresponding renormalized equation, this equation can, in addition to the trivial perturbative case, have two
qualitatively different non-trivial solutions. These solutions are standardly referred to as {\em scaling} \cite{vonSmekal:1997is,Lerche:2002ep,Zwanziger:2001kw,Alkofer:2004it,Fischer:2008uz,Fischer:2006vf,Pawlowski:2003hq} and {\em decoupling} solution \cite{Fischer:2008uz,Boucaud:2008ji,Aguilar:2008xm,Dudal:2008sp,Alkofer:2008jy}, see \cite{Cucchieri:2008fc,Sternbeck:2008mv,Bowman:2007du,Bogolubsky:2009dc,Bornyakov:2008yx} and references therein for lattice results. Note that both solutions yield a confining Polyakov-loop potential \cite{Braun:2007bx}.\\
For each of these two classes of solutions one can subsequently solve the remaining system of coupled equations, starting with the gauge sector and continuing with the matter part. 
In the scalar sector different solutions are found in the massive and the massless case. The complete fixed point structure is given in table \ref{tab:fixed-points} and presents the main result of our analysis. The mass of the scalar is here taken into account via the parameter $\eta$, thus these fixed points in principle hold for both, a massless and a massive scalar field, but in the following we will see that there are further subtleties for massive particles.
\begin{table}
\begin{tabular}{|c|c|c|c|c|c|c|c|c|c|}
\hline 
 & $\delta_{gh}$ & $\delta_{gl}$ & $\delta_{s}$ & $\delta_{gg}$ & $\delta_{3g}$ & $\delta_{4g}$ & $\delta_{sg}$ & $\delta_{sgg}$ & $\delta_{4s}$\tabularnewline
\hline 
scaling & $-\kappa$ & $2\kappa$ & $\eta$ & $0$ & $-3\kappa$ & $-4\kappa$ & $-\eta\!-\!\kappa\:\vee\:0$ & ($-\eta\!-\!2\kappa\:\vee\:0$) & ($-\eta$)\tabularnewline
\hline
decoupling  & $0$ & $1$ & $\eta$ & $0$ & $0$ & $0$ & $0$ & $0$ & $0$\tabularnewline
\hline
tree-level & $0$ & $0$ & $\eta$ & $0$ & $0$ & $0$ & $0$ & $0$ & $0$\tabularnewline
\hline
\end{tabular}
\caption{The anomalous power law exponents of the primitively divergent Green functions
for the different fixed points of the fundamentally charged scalar
theory in the uniform limit. The results are given for both massive scalars $\eta=1$ and massless scalars $\eta=0$. The given anomalous exponents represent
the scaling of the propagators of the ghosts $\delta_{gh}$, the gluons
$\delta_{gl},$ and the scalars $\delta_{s}$ as well as of the ghost-gluon
vertex $\delta_{gg}$, the 3- and 4-gluon vertices $\delta_{3g}$ and
$\delta_{4g}$, the scalar-gluon vertex $\delta_{sg}$, the scalar-2-gluon
vertex $\delta_{sgg}$ and the 4-scalar vertex $\delta_{4s}$. These
power laws are only valid up to possible logarithmic corrections. The full
vertices include also the canonical scaling dimension $-1$ for the
propagators and $\frac{1}{2}$ for the 3-point vertices. Note that the uniform limit is not sufficient to determine the dominating scaling exponents of the scalar-2-gluon and the 4-scalar vertex. The values given here in parentheses will be corrected below by the inclusion of kinematic divergences for massive scalars. The value of $\kappa$ is fixed by an explicit IR solution and the best known
value is $\kappa\!\approx\!0.59$ \cite{Zwanziger:2001kw,Lerche:2002ep,Pawlowski:2003hq}.}\label{tab:fixed-points}
\end{table}

\noindent It is a remarkable result that, as in the case of QCD \cite{Alkofer:2008tt,Schwenzer:2008vt}, the additional scalar dynamics does not change the known IR fixed
point structure of the Yang-Mills sector. The latter features two rather different possible IR scenarios of
the continuum theory. First, the decoupling solution with a massive
gluon propagator and no IR enhanced Green
functions. Second, the scaling solution with a strongly IR singular
ghost propagator, which induces similar divergences in gluonic vertex
functions but an IR-suppressed gluon propagator. The
scaling solution provides a mechanism for the confinement of the gauge
degrees of freedom within the Kugo-Ojima \cite{Kugo:1979gm} and
Gribov-Zwanziger \cite{Gribov:1977wm,Zwanziger:1991gz} scenarios.
Moreover, as discussed in detail below it also provides a mechanism for the
confinement of the matter fields \cite{Alkofer:2008tt}.
Yet, only the first of these solutions has been observed in current
(4-dimensional) lattice simulations \cite{Cucchieri:2008fc,Sternbeck:2008mv,Bowman:2007du,Bogolubsky:2009dc,Bornyakov:2008yx} and it is a challenging
question whether the IR behavior they show is indeed the only solution
that is realized when the continuum limit is taken, cf. e.g. \cite{Fischer:2008uz,Zwanziger:2009je,Maas:2009se,Kondo:2009qz,vonSmekal:2008ws,vonSmekal:2008es}.\\
Whereas the scalar sector is entirely trivial in the decoupling solution,
there are two qualitatively different IR fixed points for the scaling solution, one with trivial vertices and another one with strongly divergent vertices. These two different fixed points are analogous to the case of QCD. In the massive case the divergent solution for the full scalar-gluon vertex including its canonical dimension features in particular precisely the same IR power law $-1/2-\kappa$ as the quark-gluon vertex in QCD \cite{Alkofer:2006gz}. In both theories this IR divergence of the vertex is completely induced by the gauge sector. The above scaling laws are not altered by the neglected 2-loop diagrams, as is checked explicitly in \cite{Leo}. Furthermore, the DSEs for higher $n$-point functions with $n>4$ are linear. Accordingly these equations cannot induce additional non-trivial fixed points and therefore the infrared exponents of these $n$-point functions are purely determined by the lower $n$-point functions. Moreover, we find that in all cases there are leading graphs that do not involve 4-point functions, in complete analogy to the case of QCD where such 4-point functions are not primitively divergent. \\
In the approximation discussed so far, the scalar 4-point vertices in table \ref{tab:fixed-points} feature in the massive case only rather mild divergences. In particular, the obtained divergence of the one-particle irreducible 4-scalar vertex is less than the divergence of the corresponding one-particle reducible vertex built using a gluon exchange via two scalar-gluon vertices. We will show below that this is a shortcoming of the present restriction to uniform scaling exponents. In general a more diverse IR behavior can be realized in which Green functions have also kinematic divergences if only a subset of the external momenta vanishes  whereas the others remain finite. In this case the loop integrals can receive dominant contributions from hard modes even when all external scales are small. Correspondingly, kinematic divergences can alter the uniform power laws. Due to this the present results for the 4-scalar and the scalar-2-gluon vertex from the uniform limit in table \ref{tab:fixed-points} are given in parentheses. They are only correct for massless particles but underestimate the actual divergence in the massive case. We will explain in the next section how the correct exponents are recovered once kinematic divergences are taken into account. \\
Finally, we want to point out that a solution with a divergent scalar propagator is not possible, due to the self-interaction of the scalar, cf. \cite{Huber:2009wh} for a general treatment of this issue.

\section{Kinematic singularities and static confinement}
As mentioned before, the uniform solution discussed so far presents only a special case
of the actual possible IR behavior. In the following we will extend this by the inclusion of kinematic divergences of the vertex functions \cite{Fischer:2006vf,Alkofer:2008jy,Alkofer:2008dt}. Such kinematic divergences provide a mechanism for a long-range interaction
that can confine quarks in quenched QCD \cite{Alkofer:2008tt}. Therefore we are interested whether this mechanism can also confine static scalar sources, i.e. if the relevant Green functions feature the same IR scaling exponents. A general discussion of kinematic divergences in the case of the scalar
theory considered here is quite complicated due to the additional
interactions and the many possible kinematic limits of the 4-point
vertices, and is therefore beyond the scope of this work. However, in the following we
will present the line of argument that shows that the kinematic divergence of the scalar-gluon
vertex in the static case is indeed as strong as that of the quark-gluon vertex in QCD. To this end we restrict the discussion in the following to the quenched approximation where the scalar sector does not affect
the gauge sector and can be analyzed independently. Since it was found above that the 4-point functions do not alter the IR scaling laws we can restrict our investigation to the equation for the scalar-propagator and the scalar-gluon vertex. We have checked explicitly in \cite{Leo} that this remains true if taking into account the enhanced scaling of the vertices obtained below. \\ 
A comparison with the corresponding QCD equations shows that the equations for the scalar-propagator and the scalar-gluon vertex in fig. \ref{fig:scalar-DSEs} are diagrammatically similar to the corresponding equations for the quark-propagator and the quark-gluon vertex, except for additional terms in the scalar theory that stem from the additional 4-point interactions. In order to find the IR exponents a refined power counting analysis has to be employed that takes into account the possibility of momentum and mass scales that stay finite when the IR limit is taken. As shown in detail in \cite{Alkofer:2008jy}, the scale separation
between these soft and hard scales in the IR limit allows to identically
decompose the arising loop integrals into several integrals that depend
only on a single external scale which directly determines the scaling of the corresponding contribution. This yields a system for the IR exponents of the scalar propagator $\delta_s$ and the scalar-gluon vertex in the uniform limit $\delta_{sg}^u$ as well as $\delta_{sg}^s$ and $\delta_{sg}^{gl}$ in the limits that only a scalar respectively gluon momentum vanishes. Strikingly the additional terms from the 4-point interactions in these equations can be shown to be subleading \cite{Leo} using constraints from the inequivalent towers of DSEs and RGEs \cite{Fischer:2006vf}. Correspondingly the scalar system effectively reduces to a DSE system that is up to the different canonical dimensions identical to that of quenched QCD. The different canonical dimensions only further suppress terms that are subleading in the case of QCD. Thus the same fixed point structure is realized and the scalar-gluon vertex features the same scaling behavior as the quark-gluon vertex. \\
The solution is given in table \ref{tab:kin-fixed-points}, where the superscripts denote the soft momenta in the specific limit. It features strong kinematic divergences of the scalar-gluon vertex in the limit that
only the external gluon momentum vanishes. Note that this soft-gluon
divergence is of the same size as the uniform divergence and that
the inclusion of kinematic divergences does not change the uniform
scaling of the scalar-gluon vertex.

\begin{table*}
\begin{tabular}{|c|c|c|c|c|}
\hline 
 & $\delta_{s}$ & $\delta_{sg}^{u}$ & $\delta_{sg}^{gl}$ & $\delta_{sg}^{s}$\tabularnewline
\hline 
scaling & $\eta$ & $-\eta\!-\!\kappa\:\vee\:0$ & $-\eta\!-\!\kappa\:\vee\:0$ & $0$\tabularnewline
\hline
decoupling & $\eta$ & $0$ & $0$ & $0$\tabularnewline
\hline
\end{tabular}
\caption{The anomalous power law exponents of the leading correlation functions
of the quenched scalar model when taking into account kinematic divergences.
\label{tab:kin-fixed-points}}
\end{table*}

\begin{figure}
\parbox[c][1\totalheight]{0.5\columnwidth}{
\includegraphics[scale=0.16]{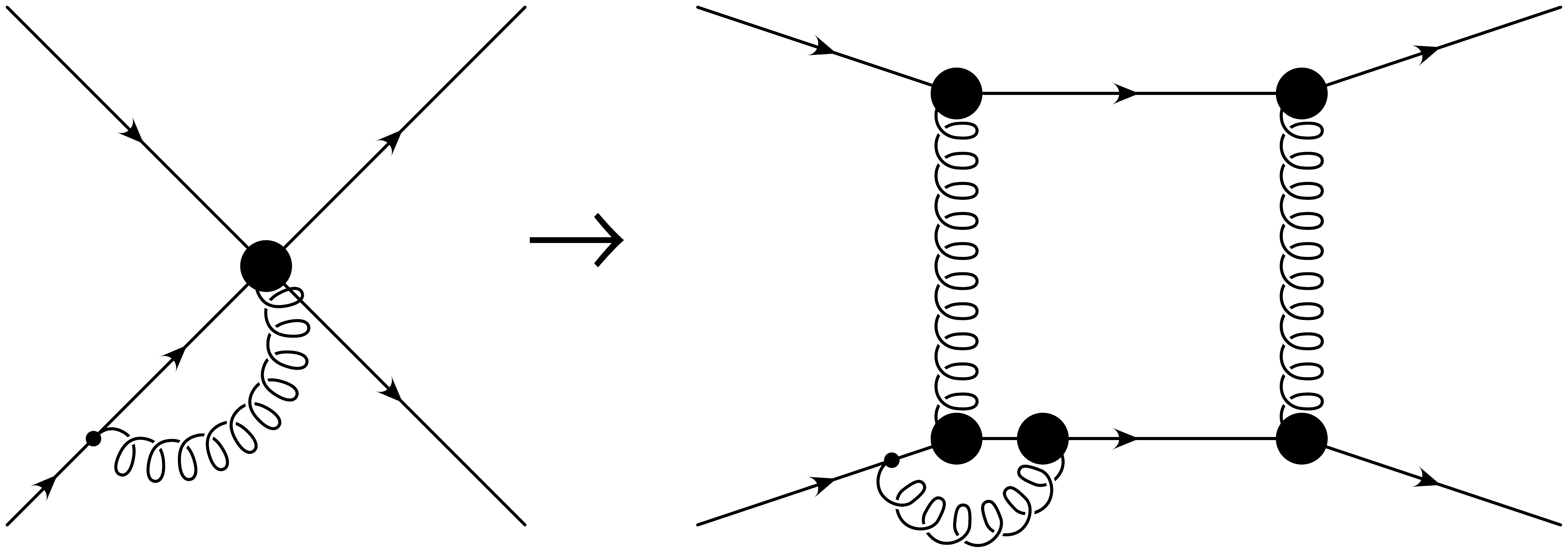}
}%
\begin{minipage}[c][1\totalheight]{0.5\columnwidth}%
\flushright\includegraphics[scale=0.17]{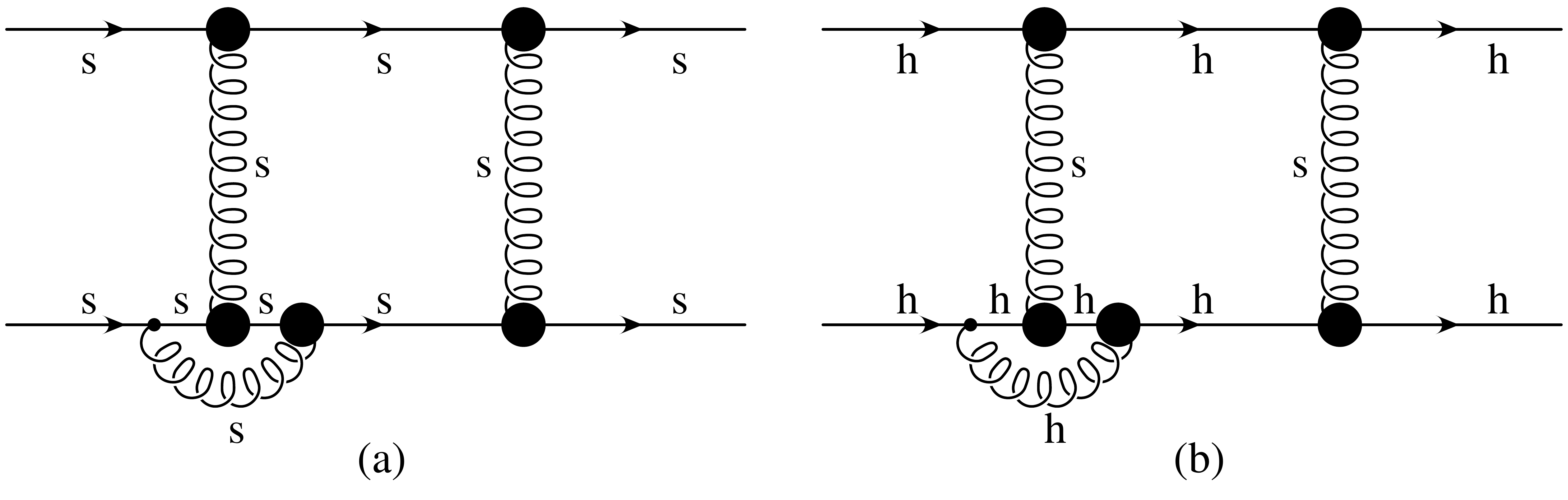}%
\end{minipage}
\vspace*{1.3cm}
\caption{\emph{left}: An IR leading contribution to the 4-scalar vertex when
inserting a corresponding contribution arising in the DSE of the 4-scalar-gluon
vertex, \emph{right}: Kinematic configurations that yield the leading
order contributions for the scaling of the vertex in the case that all external momenta vanish uniformly (a) and in the limit when only the momentum transfer between the scalars becomes small (b). The labels denoted that the momenta running through the corresponding propagators are soft $s$ and vanish in the IR limit or that they are hard $h$ of the order of the scalar mass. \label{fig:4-point}}

\end{figure}

\noindent However as in the case of QCD, for higher uniform correlation functions
the consideration of kinematic divergences is required in a DSE study
even to get the proper scaling in the uniform limit. As explained in \cite{Alkofer:2008tt} this is a peculiarity of the DSEs owing to
the fact that they involve a bare vertex in each graph. Thereby in
a theory with enhanced vertices IR strength can be missing in the
lowest order corrections and is represented dynamically in the contributions
from correlation functions with one more external leg.
To see this consider the last term in the equations of the 4-scalar
vertex that involves a 5-point function. The latter satisfies its
own DSE which contains a graph with a simple gluon exchange correction
involving only scalar-gluon vertices. Inserting this correction into
the corresponding term in the 4-scalar DSE as visualized in the left
part of fig. \ref{fig:4-point} yields a 2-loop diagram that looks
very similar to the ordinary gluon exchange graph in the 4-scalar
DSE with the difference that the bare vertex has now a vertex correction.
When the scalars are massive, this vertex correction loop receives
contributions from hard loop momenta and scales only due to the kinematic
divergence of the dressed vertex, cf. fig. \ref{fig:4-point} (a). In the heavy quark limit the hard
loop simply shrinks to a point and precisely presents a forth dressed
vertex which is present from the outset in functional RG equations.
This mechanism finally yields the leading IR scaling laws for the
4-point vertices in table \ref{tab:fixed-points}.\\
Now let us discuss the interaction between static scalar sources which
is described by the heavy mass limit of the full 4-scalar vertex.
Here the relevant limit it is not the uniform kinematic configuration
where momenta of the external scalars are in the IR regime and the
scalars would correspondingly be on the light cone in Minkowski space,
but the exactly opposite limit where their mass is large and the external
momenta are of the order of this large scale. Nevertheless, when the scalars
are far spatially separated the exchanged gluon momentum becomes soft
and the correlator can be IR enhanced. Due to the same mechanism described
above the leading contribution arises from the graph on the left of
fig. \ref{fig:4-point} and the dominant kinematic contribution is
given by graph (b) on the right hand side - again effectively adding a forth dressed
vertex. The corresponding graph with a soft gluon exchange scales
therefore in the IR as
\begin{equation}
(p^{2})^{2}\Bigl((p^{2})^{-\frac{1}{2}-\kappa}\Bigr)^{4}\Bigl((p^{2})^{-1+2\kappa}\Bigr)^{2}=(p^{2})^{-2}
\end{equation}
which leads after Fourier transformation in the static limit to a
linear potential in coordinate space
\begin{equation}
V(r)\sim\int d^{3}p\frac{e^{ipr}}{p^{4}}\sim|r|\:.
\end{equation}
Correspondingly, scalars are subject to the same static confinement
mechanism as quarks in the case of QCD. In contrast the decoupling solution
does not provide a corresponding mechanism, i.e. if the decoupling solution were confining this would not be reflected in any $n$-point function with finite $n$. \\
It is easy to convince oneself that the universality of the long-range interaction between static fundamental color sources in the discussed mechanism is not restricted to the explicitly studied cases of Dirac fermions and scalars but should actually hold for matter fields in any representation of the Lorentz group. This follows since the performed IR analysis is totally independent of the Lorentz structure and depends only on the topology of the individual graphs in the DSEs and the involved propagators and vertices. The case of a complex scalar field considered here presents the renormalizable theory of matter fields coupled to a Yang-Mills sector with the most general interactions in four spacetime dimensions. The explicitly considered cases of scalars and Dirac fermions present precisely the two distinct possibilities for the dynamics as far as the topology of possible graphs is concerned. Deviations from the scaling laws obtained within a power counting analysis are only possible if there are identical cancellations of the leading graphs in the DSEs. In the case of the scalar- respectively quark-gluon vertex there is actually only a single leading diagram in the corresponding DSE, cf. figs. \ref{fig:scalar-DSEs} and \ref{fig:gh-box}, so that cancellations between different graphs are impossible here. This shows that the universality of the long-range interaction between static fundamentally charged sources, which is a natural property in the lattice framework, is indeed realized in the functional framework as well. \\
Finally, we note that although in this work we studied only gauge dependent Green functions, the above 4-point correlator represents the lowest term in a power series representation of the exponential arising in the corresponding gauge invariant correlator where the two quark sources are connected by a Wilson line. Therefore, in case this leading term is not cancelled identically by higher order terms in the series, this gauge independent quantity likewise shows the observed long-range interaction.

\section{Conclusion}

We have studied the IR fixed point structure of a non-Abelian gauge
theory coupled to a scalar matter field in the fundamental representation
of the gauge group as a schematic model for the QCD dynamics. We find that the
IR fixed point structure is indeed identical to the case of QCD and
that for one type of solutions a kinematic divergence of the scalar-gluon vertex induces a linear
confining interaction between static sources. The qualitatively identical
confinement aspects of the scalar model compared to QCD show that
this confinement mechanism is indeed universal and does not depend
on the particular features of the matter fields. Instead the long-range interaction between fundamental sources is a property of the gauge sector in complete analogy to results of lattice gauge simulations.
Therefore, it is merely a technical complication, that within functional
Green function methods the coupling of a fundamental color source
to the gauge sector has to be obtained from the static limit of the
dynamical equations of the corresponding matter fields. In this limit
the kinematic divergence of the matter-gauge vertex simply describes
the non-trivial dressing of the static color source. \\
Finally, we want to emphasize that due to the presented results
the scalar theory presents an ideal model system to study a potential confinement
mechanism in detail. Within functional approaches the minimal system
of coupled integral equations for the matter sector decreases from
14 for fermionic fields to 3 in the scalar case. This should strongly
simplify the numerical treatment and allow to study quantitative aspects
of the mechanism. Even more importantly, the dramatic simplifications of scalars
compared to fermions in lattice gauge simulations could allow to test
this mechanism in lattice simulations.

\acknowledgments
We are grateful to Christian S. Fischer, Jeff Greensite, Markus Huber, Axel Maas, Stefan Olejnik and Jan M. Pawlowski for helpful discussions, furthermore we thank Markus Huber and Jan M. Pawlowski for a critical reading of the manuscript. LF acknowledges financial support by the Helmholtz-Alliance HA216/EMMI.

\end{document}